\documentclass[twocolumn,superscriptaddress,showpacs,pra]{revtex4}
\usepackage{epsfig}
\usepackage{delarray}
\usepackage{amsmath, amssymb}
\usepackage{bm}
\newcommand{\ket}[1]{| \, #1 \rangle}

\begin{document}
\title{Enhancing the W State Quantum Network Fusion Process  with A Single Fredkin Gate}

\author{Sinan Bugu}
\affiliation{Department of Computer Engineering, Okan University, Istanbul, Turkey}

\author{Can Yesilyurt}
\affiliation{Department of Computer Engineering, Okan University, Istanbul, Turkey}

\author{Fatih Ozaydin}
\email{MansurSah@gmail.com}
\affiliation{Department of Computer Engineering, Okan University, Istanbul, Turkey}

\begin{abstract}
Integrating a single Fredkin (controlled swap) gate to the previously introduced W state fusion mechanism \emph{ (Ozdemir et al, N. J. Phys. 13, 103003, 2011)} and using an ancillary photon, we increase the size of the fused W states and essentially, we improve the success probability of the fusion process in a promising way for a possible deterministic W state fusion mechanism. Besides fusing arbitrary size W states, our setup can also fuse Bell states to create W states with a success probability $3/4$ which is much higher than the previous works. Therefore using only this setup, it is now possible to start with Bell pairs to create and expand arbitrary size W states. Since higher probability of success implies a lower cost of resource in terms of the number of the states spent to achieve a target size, our setup gives rise to more cost-efficient scenarios.

\end{abstract}
\maketitle

When the number of particles forming an entangled state increases beyond two (i.e., two corresponding the bipartite case), a variety of states with more complex and different entanglement structures emerge. More interestingly, these states fall in inequivalent classes with Greenberger-Horne-Zeilinger (GHZ), W, Dicke and cluster states being the well-known examples. States belonging to different classes cannot be converted to each other even under stochastic local operations and classical communications (SLOCC) \cite{NJPRef4}.  Understanding the entanglement structures and the formation of states belonging to different inequivalent classes is important not only for the general entanglement theory, but also for their vital roles in various quantum information processing tasks such as some quantum algorithms, quantum key distribution, quantum teleportation, measurement based quantum computation, etc. It is known that some states are more suitable for specific tasks than the others \cite{NJPRef5,NJPRef6,NJPRef7, NJPRef8, NJPRef9, NJPRef10,NJPRef11,NJPRef12}. Thus, preparation of task-specific multipartite entangled states could benefit the quantum information science significantly. However, it is also crucial that these states are prepared using the resources efficiently with minimal costs. Therefore, simple and efficient schemes and methodologies to prepare large-scale multipartite entangled states are being sought, and there have been tremendous efforts put into this endeavour.

Bipartite entangled states are understood very well.  In principle, starting with EPR pairs, we can prepare arbitrary bipartite entangled states. We now know how to prepare, characterize, manipulate and use bipartite entangled states for specific tasks. We also know how to use EPR pairs as resources to prepare multipartite entangled states such as GHZ, W and cluster states \cite{NJPRef17,NJPRef18,NJPRef19,NJPRef20,NJPRef21,NJPRef22,NJPRef23,NJPRef24,NJPRef25}. However, despite the great efforts the theory and experiments on multipartite entanglement have been lagging. In the last decade, expansion and fusion operations are proposed and demonstrated as efficient ways of preparing large scale multipartite entangled states. In the expansion operation, the number of qubits in an entangled state is increased by one or two qubits at a time by locally accessing only a limited number of qubits of the original state. Fusion operation, on the other hand, prepares a larger entangled state by fusing two or more multipartite entangled states with the condition that access is granted only to one qubit of each of the states entering the fusion operation. Expansion and fusion operations have been demonstrated experimentally for GHZ and cluster states. For W-states, on the other hand, experiments have shown the possibility of efficient expansion of a seed W-state by one or two qubits at a time. Although there is a theoretical proposal, fusion operation for W-states has not been experimentally demonstrated yet.

Currently, efficient preparation and expansion of GHZ and cluster states are well-known \cite{NJPRef13, NJPRef14}; however, this is not the case for W states. Among many proposals \cite{NJPRef31,NJPRef32,NJPRef26} the best setup known for fusing W states was proposed by Ozdemir et al \cite{Sahin-NJP-Oct}. However, as the authors stated there is still room for improvement of the fusion mechanism to achieve more efficient preparation of larger W-states. In this paper, we will show how one can improve the efficiency of the fusion gate proposed by Ozdemir et al \cite{Sahin-NJP-Oct} by integrating a Fredkin gate and using ancillary photons.

\begin{figure}[t]
  \centering
      \includegraphics[width=0.25\textwidth]{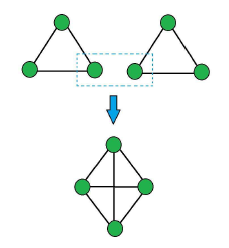}
  \caption{(color online). The fusion process of \cite{Sahin-NJP-Oct}. One photon from the W states of each party is sent to the fusion mechanism (indicated in a dashed blue rectangle), resulting a larger W state.}
\end{figure}

In the fusion process, two parties, Alice and Bob, possess $n$- and $m$-partite polarization encoded W-states, $|W_n\rangle_A$ and $|W_m\rangle_B$, respectively, and they wish to fuse their states to obtain a larger W state. The concept of fusion operation is depicted in Fig.1. Following a similar notation to that of \cite{Sahin-NJP-Oct}, we denote the polarization entangled states of Alice and Bob as
\begin{equation}
|W_n\rangle_{A} = {1 \over \sqrt{n} } ( | (n-1) _H \rangle_a | 1_V \rangle _1  + \sqrt{n-1} | W_{n-1}  \rangle_a |1_H \rangle _1 )
\end{equation}
\begin{equation}
|W_m\rangle_{B} = {1 \over \sqrt{m} } ( | (m-1) _H \rangle_b | 1_V \rangle _2  + \sqrt{m-1} | W_{m-1}  \rangle_b |1_H \rangle _2 ).
\end{equation}
where photons in modes 1 (2) are sent to a fusion gate and the photons in mode a (b) are kept intact at their site. In this notation a tri-partite W-state is written as $|W_3\rangle_A = {1 \over \sqrt{3} }(|HHV\rangle_A+|HVH\rangle_A+|VHH\rangle_A)={1 \over \sqrt{3} } ( | 2 _H \rangle_a | 1_V \rangle _1  + \sqrt{2} | W_{2}  \rangle_a |1_H \rangle _1 )$ with $W_{2}$ corresponding to the EPR pair $W_{2}= \frac{1}{\sqrt{2}}(|HV\rangle+|VH\rangle)$.

\begin{figure}[!b]
  \centering
      \includegraphics[width=0.4\textwidth]{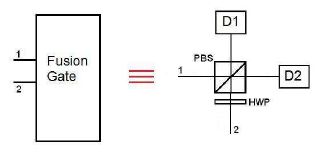}
  \caption{ (color online). Fusion Gate of \cite{Sahin-NJP-Oct}, (which we call FG from now on, as a single gate). D1 and D2 each include one QWP, one PBS and two photon detectors. }
\end{figure}
The fusion gate, as shown in Fig.2, accepts one photon from each of Alice and Bob and rotates the polarization of one of the photons by $\pi/2$. The photons are then mixed on a polarizing beamsplitter (PBS) whose output modes are measured in $\{\ket{D},\ket{\bar{D}}\}$ basis where $\ket{D}=(\ket{H}+\ket{V})/\sqrt{2}$ and $\ket{\bar{D}}=(\ket{H}-\ket{V})/\sqrt{2}$ with the detectors D1 and D2. Whenever the  photons in modes 1 and 2 have orthogonal polarizations, a coincidence detection takes place between the detectors D1 and D2. If the photons have the same polarization then the two photons go either to D1 or to D2, and no coincidence takes place. There are thus four possible cases: (i) When both photons are H-polarized, only D1 {\it clicks}, implying that the fusion operation has failed, and the remaining photons at the sites of Alice and Bob becomes $|W_{n-1}\rangle_{A}$ and $|W_{n-1}\rangle_{B}$, i.e., still W-states but with reduced number of qubits. As noted by Ozdemir et al \cite{Sahin-NJP-Oct}, the remaining W-states can be re-sent to the fusion gate for a second attempt of fusing. (ii) When both photons are V-polarized,  only D2 {\it clicks}. This is the failure case, too, and since all the V-polarized photons are now destroyed during the detection, remaining photons at the sites of Alice and Bob are all H-polarized. Thus the initial W-states are destroyed. (iii) When photon in mode 1 is H-polarized and that in mode 2 is V-polarized, photons entering the PBS has the same H-polarization after the polarization rotation in mode 2. Thus, each of the detectors will receive one H-polarized photon, and a coincidence will be observed. (iv) When photon in mode 1 is V-polarized and that in mode 2 is H-polarized, photons entering the PBS has the same V-polarization after the polarization rotation in mode 2. In this case, too, a coincidence will be observed as each detector will receive one V-polarized photon. Since the detection is performed in $\{\ket{D},\ket{\bar{D}}\}$, the cases (iii) and (iv) are indistinguishable. The superposition of the states observed for (iii) and (iv) then yields a larger W-state as a result of successful fusion operation. The final W-state upon successful fusion is $|W_{n+m-2}\rangle$, where $-2$ is due to the destroyed photons in the detection process. Table I depicts these four possible cases and their probabilities. It is seen that the success probability of the fusion gate is
\begin{equation} P^{FG}_s = { n+m-2 \over nm }  \end{equation}
where the subscript FG denotes fusion gate.

\begin{figure}[!t]
  \centering
      \includegraphics[width=0.45\textwidth]{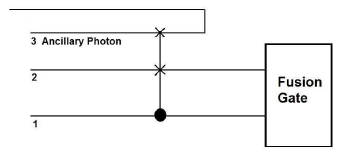}
  \caption{(color online). Integrating a Fredkin gate to FG. The ancillary photon is returned back to the (fused) state, therefore increasing the size of the state (in the success case).}
\end{figure}

\begin{table}[b]
\centering
\begin{tabular}{c c c}
\hline
Input & Probability & Result  \\ 
\hline \\
H, H & ${ (n-1)(m-1)  \over nm  }$  & Recycle \\[1ex]
H, V &${ (n-1)  \over nm  }$ 	& Success \\[1ex]
V, H  &${ (m-1)  \over nm  }$ 	& Success \\[1ex]
V, V  & ${ 1  \over nm  }$	& Failure \\ [1ex] 
\hline
\end{tabular}

\label{table:cases} 
\caption{ Four possible cases of FG.}
\end{table}

Various scenarios and strategies have been considered to decrease the cost of preparing larger W-states by fusing W-states \cite{Sahin-NJP-Oct}. None of the studied strategies are claimed to be optimal, therefore one wonders whether there exists an optimal strategy which can prepare arbitrarily large W-states by fusion processes and whether one can improve the success probability of fusion gate by modifying the basic fusion gate set-up of  Ozdemir et al \cite{Sahin-NJP-Oct}, therefore achieving a lower cost. In this work, we tackle the latter and show that by integrating a Fredkin gate to input ports of the basic fusion gate (FG) of \cite{Sahin-NJP-Oct} and using an ancillary photon of H polarization, as illustrated in Figure 3, we can improve the success probability. A good point of this new scheme is that the ancillary photon is not consumed but added to the resultant W-state.

\begin{figure}[!t]
  \centering
      \includegraphics[width=0.35\textwidth]{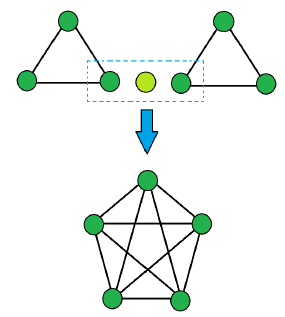}
  \caption{(color online). The fusion presented herein. One photon from the W states of each party is sent to the fusion mechanism which includes a Fredkin gate, together with an ancillary photon (light green).}
\end{figure}

A Fredkin gate, also known as controlled-SWAP gate, is a universal gate for quantum computing, that is any logical or arithmetic operation can be constructed using only Fredkin gates and ancillary qubits. The Fredkin gate is a three-qubit gate which swaps the target qubits if the first qubit is logical one. In Fig. 3, a V-polarized photon in mode 1 (control qubit) will swap the target qubits in modes 2 and 3. In our scheme, the qubit in mode 3 is defined by an H-polarized photon. Thus, we have 4-cases out of 8 possible inputs for the three-input gate, i.e., $|HHH\rangle_{123}$, $|HVH\rangle_{123}$, $|VHH\rangle_{123}$ and $|VVH\rangle_{123}$. Then the action of the Fredking gate on these input states will lead to $|HHH\rangle_{123}\rightarrow|HHH\rangle_{123}$, $|HVH\rangle_{123}\rightarrow|HVH\rangle_{123}$, $|VHH\rangle_{123}\rightarrow|VHH\rangle_{123}$ and $|VVH\rangle_{123}\rightarrow|VHV\rangle_{123}$. Assuming that the photons in modes 1 and 2 are coming from the W states of Alice and Bob, the action of the Fredkin gate is just to exchange a V-photon coming from Bob's W state with the ancillary H-photon if the photon coming from Alice's W-state is also V-polarized. In all the other cases, there is no exchange process. As we have discussed above the fusion gate of Ozdemir et al \cite{Sahin-NJP-Oct} fails when the photons coming from Alice and Bob are  V-photons.\\

\begin{table}[!t]
\centering
\begin{tabular}{c c c c}
Input 	& Probability 	& FG 	& FG\&F \\ 
\hline \\
H, H 	& ${ (n-1)(m-1)  \over nm  }$  	& Recycle 	& Recycle\\[1ex]
H, V 	& ${ (n-1)  \over nm  }$ 		& Success 	& Success\\[1ex]
V, H  	& ${ (m-1)  \over nm  }$ 		& Success 	& Success \\[1ex]
V, V  	& ${ 1  \over nm  }$		& Failure 	& Success \\ [1ex] 
\hline
\end{tabular}
\label{table:cases} 
\caption{The enhancement due to Fredkin gate and ancillary photon. The failure case is turned to a success case.}
\end{table}

 Through the integration of the Fredkin gate, we see that even if the photons coming from Alice and Bob are V-polarized $|VV\rangle_{12}$, we will have $|VH\rangle_{12}$ at the input of the fusion gate. Polarization rotation at port 2 of the PBS, then leads to $|VV\rangle_{12}$. Since now the polarization of the photons at the PBS inputs are the same, the photons will go to different output ports, leading to coincidence detection. The combined action of the fusion gate and Fredkin gate leads to coincidence detection (i.e., successful fusion events) for all cases except for the case when the photons coming from Alice and Bob are H-polarized photons. Thus, three out of four possible cases lead to successful fusion. The overall state of the photons remained intact at Alice's and Bob's sites together with the ancillary qubit is a W-state with $n+m-1$ photons, i.e., $|W_{n+m-1}\rangle$.\\

There are \textbf{three} main improvements over the scheme of Ozdemir et al \cite{Sahin-NJP-Oct}. First, thanks to the Fredkin gate and ancillary state, the success probability is increased to

\begin{equation} P^{FG\& F}_s = { n+m-2 \over nm } + {1 \over nm } = { n+m-1 \over nm }. \end{equation}

\noindent This is because one of the failure cases becomes a success with the Fredkin gate (see Table II).  Second, the final W-state prepared upon the successful operation of the gate has one more qubit than that of the scheme of Ozdemir et al \cite{Sahin-NJP-Oct}, because the ancillary qubit after the action of the Fredkin and fusion gates are added to the final state.Third, not only arbitrary size W states, $|W_n\rangle$ and $|W_m\rangle,$ $n \geq 3$ and $m\geq3$; can be fused but using setup presented herein, W states of $n=2$ and/or $m=2$ (which are Bell states) can also be fused.

The success probability of our setup to create a $|W_3\rangle$ state from a single (ancillary) photon and two Bell states is $3/4$ which is much higher than the previous setups, i.e. the probability of success to create $|W_3\rangle$ states in Ref.\cite{NJPRef31} from a single photon and a Fock state is $3/16$; in Ref.\cite{NJPRef32} from a single photon and a Bell pair is $3/10$; and in Ref.\cite{NJPRef26} (experimentally) from two Bell states is $3/27$.

Besides the higher success probability, an important advantage of our setup appears to be that in order to create large scale W-state networks, using only this setup, one can start with single photons and Bell pairs and continue expanding the network to any large size, whereas using previous setups, one should first use the setup of one of \cite{NJPRef31,NJPRef32,NJPRef26} to create $|W_3\rangle$ states and then send these $|W_3\rangle$ states to the setup of \cite{Sahin-NJP-Oct} to expand the network. In case of recycle, the scenario is the reverse, i.e. during the expanding process of \cite{Sahin-NJP-Oct}, whenever the size of the $|W\rangle$ states decreases to $2$, these Bell states should either be left (taking new $|W_3\rangle$ states) or sent to the setup of one of \cite{NJPRef31,NJPRef32,NJPRef26} to obtain $|W_3\rangle$ states again.

Since there are proposals on implementing Fredkin gate using linear optical elements \cite{Fiurasek-Fredkin}, we do believe that with the pace of developments in quantum optical technologies, we are not far away from implementing W-state fusion gates with integrated Fredkin gates. On the other hand, in this paper we have not taken into account the effects of losses, inefficiencies of the optical gates, photon sources and the detectors, as well as the memory issues. The effects of the deviations from the ideality on the performance of the proposed fusion gate can be taken into account using the methods and techniques developed in Refs. \cite{NJPRef32,Varnava,ozdemir1,Gong,Wei,ozdemir2,Kieling3}. Besides many quantum information tasks requiring large scale quantum networks, the endeavor on efficient creation of such networks is also important for improving the understanding of percolation behavior and the quantum critical phenomena in quantum networks with different network topologies and connectivity provided by the shared entanglement \cite{NatPhysPercolation,Kieling1,Kieling2}. We believe that a further study on these issues would shed light onto the scalability of not only this approach but also of any mechanism devoted to expansion and fusion of multipartite entanglement networks.

In conclusion, integrating a Fredkin (controlled swap) gate to a previously proposed W state fusion gate, together with the idea of using an ancillary photon, not only we have increased the size of the fused state but also, we have increased the total success probability of the fusion process. We also managed to fuse a Bell state either with another Bell state or an arbitrary size W state. Since the cost of expanding W class quantum networks is defined in terms of the resource spent over the probability of success; the setup we propose in this work, decreases the cost of expanding W class quantum networks.


\end{document}